\documentclass[12pt,epsf]{article}
\usepackage{graphicx}
\usepackage{lscape}
\usepackage{color}
\usepackage{slashbox}
\usepackage{comment}
\usepackage{slashed}
\begin{document}

\def\a{\alpha}
\def\b{\beta}
\def\c{\varepsilon}
\def\d{\delta}
\def\e{\epsilon}
\def\f{\phi}
\def\g{\gamma}
\def\h{\theta}
\def\k{\kappa}
\def\l{\lambda}
\def\m{\mu}
\def\n{\nu}
\def\p{\psi}
\def\q{\partial}
\def\r{\rho}
\def\s{\sigma}
\def\t{\tau}
\def\u{\upsilon}
\def\v{\varphi}
\def\w{\omega}
\def\x{\xi}
\def\y{\eta}
\def\z{\zeta}
\def\D{\Delta}
\def\G{\Gamma}
\def\H{\Theta}
\def\L{\Lambda}
\def\F{\Phi}
\def\P{\Psi}
\def\S{\Sigma}

\def\o{\over}
\def\beq{\begin{eqnarray}}
\def\eeq{\end{eqnarray}}
\newcommand{\gsim}{ \mathop{}_{\textstyle \sim}^{\textstyle >} }
\newcommand{\lsim}{ \mathop{}_{\textstyle \sim}^{\textstyle <} }
\newcommand{\vev}[1]{ \left\langle {#1} \right\rangle }
\newcommand{\bra}[1]{ \langle {#1} | }
\newcommand{\ket}[1]{ | {#1} \rangle }
\newcommand{\EV}{ {\rm eV} }
\newcommand{\KEV}{ {\rm keV} }
\newcommand{\MEV}{ {\rm MeV} }
\newcommand{\GEV}{ {\rm GeV} }
\newcommand{\TEV}{ {\rm TeV} }
\def\diag{\mathop{\rm diag}\nolimits}
\def\Spin{\mathop{\rm Spin}}
\def\SO{\mathop{\rm SO}}
\def\O{\mathop{\rm O}}
\def\SU{\mathop{\rm SU}}
\def\U{\mathop{\rm U}}
\def\Sp{\mathop{\rm Sp}}
\def\SL{\mathop{\rm SL}}
\def\tr{\mathop{\rm tr}}

\def\IJMP{Int.~J.~Mod.~Phys. }
\def\MPL{Mod.~Phys.~Lett. }
\def\NP{Nucl.~Phys. }
\def\PL{Phys.~Lett. }
\def\PR{Phys.~Rev. }
\def\PRL{Phys.~Rev.~Lett. }
\def\PTP{Prog.~Theor.~Phys. }
\def\ZP{Z.~Phys. }


\baselineskip 0.7cm

\begin{titlepage}

\begin{flushright}
IPMU-11-0121\\
UT-11-23\\
KEK-TH-1478\\
\end{flushright}

\vskip 1.35cm
\begin{center}
{\large \bf
LHC Test of CDF $Wjj$ anomaly
}
\vskip 1.2cm
Keisuke Harigaya$^{1,2}$, Ryosuke Sato$^{1,2}$, and Satoshi Shirai$^3$
\vskip 0.4cm

{\it
$^1$Institute for the Physics and Mathematics of the Universe, University of Tokyo, \\
Kashiwa 277-8568, Japan\\
  
$^2$Department of Physics, University of Tokyo, \\ 
Tokyo 113-0033, Japan\\

$^3$ Institute of Particle and Nuclear Studies,\\
High Energy Accelerator Research Organization (KEK)\\
Tsukuba 305-0801, Japan\\
}

\vskip 1.5cm

\abstract{
We discuss a test of the CDF dijet anomaly at the LHC.
The recent observed dijet mass peak at the CDF is well fitted by a new particle with a mass of
around 150 GeV, which decays into two jets.
In this paper, we focus on only $Wjj$ signal to avoid model dependence,
and comprehensively study the LHC discovery/exclusion reach.
We found almost all the models are inconsistent with the result of the LHC,
unless only valence quarks contribute the new process.
We also discuss further prospects of the LHC search for this anomaly.
}
\end{center}
\end{titlepage}

\setcounter{page}{2}
\section{Introduction}
Recently, the CDF collaboration reported an anomaly in $Wjj$ events
\cite{Aaltonen:2011mk}.
This result indicates a new particle $X$ with mass around 150 GeV and
the production cross section associated
with a $W$ boson is around 4 pb at the Tevatron.
There are many models proposed to explain this anomaly~
\cite{Buckley:2011vc,Kilic:2011sr,Isidori:2011dp,Anchordoqui:2010zs,He:2011ss}.
Since the D0 collaboration reports result inconsistent with the CDF~\cite{Abazov:2011af},
it is very important to test the $Wjj$ anomaly at the LHC.

Each model can have individual manner to be tested.
Some model may predict the collider signals other than $Wjj$.
For example, Ref. \cite{Eichten:2011xd} discusses the testing at the LHC for the Technicolor model by using various modes.
However, the discussion using modes other than $Wjj$ is strongly model dependent.
In this paper, we focus on only $Wjj$ signal at the LHC to test various models with model independent manner, regardless of such other considerations.
We discuss possible effective interactions and resonant particles which give the $Wjj$ signal.
Then, our study does not depend on the detail of the model
and we can cover almost all the model in other references.
When we discuss the various models,
we do not care about the flavor symmetry violation or the deviation from the electroweak precision test.
To discuss them, we have to specify the whole of the model.
Furthermore, some other particles unrelated to $Wjj$ signal may compensate such a violation,
then, the comprehensive study is difficult.
We study models which can realize $Wjj$ signal at the Tevatron,
and discussed the expected signal cross section at the LHC.
We found that almost all models can be discovered or excluded at the LHC.

\section{Setup}

A lot of models proposed to explain the CDF dijet anomaly assume tree level s-channel or t-channel process\footnote{
There are some models which can not be classified to such a class.
For example, in some models~\cite{Isidori:2011dp}, new particles are pair-produced at first,
and a $W$ boson and $jj$ are generated from each new particle decay.
}
whose final state is a $W$ boson and a new particle $X$  with a mass of around 150 GeV.
In this paper, we consider the following two cases:

\begin{description}
\item[Case 1:]
{\bf $W$ boson and $X$ are produced without $WX$ invariant mass peak.}\\
In this case, to test the models with model independent manners, we concentrate on the effective theory.
We consider the cases in which the particle $X$ is described by scalar, spinor and vector field. 
We have considered possible effective operators to provide a process $p\bar p\to WX$ up to dimension 5.
In constructing the effective operators, we have respected the
$SU(3)_C\times SU(2)_L \times U(1)_{Y}$ gauge symmetry and the Lorentz
symmetry but allowed the operators which can arise by the standard model
higgs condensation.

\item[Case 2:]
{\bf another particle $Y$ is produced by s-channel then it decays into a $W$ boson and $X$.}\\
A resonant behavior of $Wjj$ signal is constrained from $m_{\ell\nu jj}$ distribution, 
but $m_{\ell\nu jj}$ peak around 280 GeV is less constrained \cite{CDF_lvjj}.
So, we can consider another particle $Y$ with the mass of 280 GeV,
and $WX$ signal is generated by the following process:
\begin{eqnarray}
p + \bar p \to Y \to W+X.\nonumber
\end{eqnarray}
We have concentrated on the case in which the narrow width approximation is valid.
In this way, we can give a model-independent result.
\end{description}

We also assumed that the CP invariance is preserved at the level of effective operators.
That is, when an effective operator is introduced, its hermite conjugate operator is also introduced.
\section{Constrains from hadron collider experiments}

\subsection{LHC}

Here, we discuss the discovery/exclusion reach at the LHC 7 TeV run.
The ATLAS group shows dijet invariant mass $M_{jj}$ distribution associated with a $W$ boson with 1.02 fb$^{-1}$ data~\cite{ATLAS_wjj}.
This data is consistent with the standard model background,
then, it gives a severe constraint on the $WX$ cross section at the LHC.
The ATLAS groups shows the data with third jet veto and the data without veto.
The data without veto gives more severe constraint on the cross section,
because third jet veto significantly decreases signal acceptance while the background can not be so reduced.
In the following of this section, we use the data without third jet veto.

When the decay width of $X$ is small enough compared to the jet resolution,
$WX$ events may give significant contribution to $M_{jj}$ distribution around 150 GeV.
We estimate signal acceptance for the event cut used in Ref. \cite{ATLAS_wjj} by using MadGraph-Pythia-PGS package \cite{Alwall:2007st}. 
Given signal acceptance, we can estimate the upper bound of $WX$ cross section by simple event number counting.
We estimate the upper bound of $WX$ cross section at 95\% C.L. by using
the RooStats tools \cite{Moneta:2010pm}.
We can get the upper bound as 16.7 pb from the present ATLAS data.\footnote{
We fixed signal acceptance when we estimate the bound of the cross section.
In fact, signal acceptance slightly depends on the particle property and
its interaction.
Roughly, its dependence is a few ten percent, which can be ignored in
our conservative estimation of the upper bound on $R_{{\rm LHC}}$.
}

In the following of this paper, we discuss the LHC discovery/exclusion by using enhancement factor $R_{\rm LHC}$,
which is defined by the ratio of cross section between at the LHC 7 TeV run and
at the Tevatron, such as,
\begin{eqnarray}
R_{\rm LHC} = \frac{\s(pp\to WX; \sqrt{s}~=~7~\TEV)}{\s(p\bar p \to WX;~\sqrt{s}~=~1.96~\TEV)}. \label{eq:ratio}
\end{eqnarray}
By using the upper bound of the cross section,
we can get the upperbound of $R_{\rm LHC}$.
To estimate this bound conservatively, we take the cross section 2 pb at the Tevatron in Eq. (\ref{eq:ratio}).
Then, $R_{\rm LHC}$ have to be lower than 8.4.
This value gives stringent constraint on the model which gives $Wjj$ signal.
The more integrated luminosity and the less systematic error can give
more severe upper bound.
The result is given by Table \ref{table:sig}.
\begin{table}
\begin{center}
\caption{
The current and expected upper bounds of $WX$ cross section at 95\% C.L. in the LHC 7 TeV run.
The row shows each integrated luminosity.
The most bottom row shows the limit when statistical error can be neglect.
The column shows the size of systematic error.
In the left column we take the same systematic error as Ref. \cite{ATLAS_wjj},
In the middle column we take the half of the systematic error as Ref.
\cite{ATLAS_wjj},
In the right column we take only statistical error.
}
\begin{tabular}{|l||c|c|c|}
\hline
& sys.err. = Ref. \cite{ATLAS_wjj} & sys.err. = $\frac{1}{2}$ Ref.
\cite{ATLAS_wjj} & only stat.err. \\
\hline
\hline
1.02 fb$^{-1}$ & 16.7 pb & 12.9 pb & 11.1 pb\\
\hline
1.02 fb$^{-1} \times 2$ & 15.9 pb & 11.6 pb & 9.3 pb\\
\hline
1.02 fb$^{-1} \times 4$ & 15.4 pb & 10.8 pb & 7.9 pb\\
\hline
1.02 fb$^{-1} \times \infty$ & 15.0 pb & 10.0 pb & \\
\hline
\end{tabular}
\label{table:sig}
\end{center}
\end{table}

\subsection{${\rm Sp\bar{p}S}$}
\label{sec: SppS}
Some model predicts the single production of $X$ by the quark or gluon
fusion at hadron colliders.  Due to the low energy kinematical cut, low energy
hadron collider can set good constraints on the production cross
section of $X$ decaying into dijet.  We utilized the results of the UA2
collaboration \cite{Alitti:1993pn}, which give the constraints on the production cross section
times the branching ratio to dijet of excited vector boson and quark.
There are two things to be noticed.

First, the sensitivity of dijet search depends on the decay width of the parent particle,
because a broad width weakens the constraint from the dijet search.
The UA2 analysis assumed the particular value for the decay width of the produced particle.
However, the assumed width is smaller than the mass resolution,
which is about 10 \% of the mass of the parent particle.  Therefore, we can 
apply their constraints on the production cross section if the decay
width is smaller than $150\times 0.1=15$ GeV.
To be conservative, we do not apply the UA2 constraint to
the particle whose decay width is larger than 15 GeV.

Second, the sensitivity also depends on
the decay mode of the parent particle.
A heavy quark jet (i.e, $c$ and $b$ quark) loses its energy
due to the decay of the heavy quark.
This results in the existence of low energy tail in the dijet mass distribution,
which lowers the sensitivity of dijet peak search.
Therefore, the upperbound of the cross section becomes loose
if $X$ decays into a heavy quark.
However, as we will see in the following of this paper,
$X$ coupling to a heavy quark is severely constrained at the LHC.
Therefore, if we concentrate on a model which does not excluded by the LHC,
we can adopt the constraint for
the particle only decay into light quarks or gluon.
The upper limit for the production cross section of $X$ is 80 pb at 90 \% C.L.
In the same way as the constraint from the ATLAS result,
we discuss the exclusion/discovery by using the ratio of the cross section at the Sp$\rm\bar p$S and the Tevatron.
We defined the ratio of cross section, such as,
\begin{eqnarray}
R_{\rm SppS} = \frac{ \s( p\bar p\to X;~\sqrt{s}~=~540~\GEV) }{ \s( p\bar p \to WX;~\sqrt{s}~=~1.96~\TEV ) }. \label{eq:ratiospps}
\end{eqnarray}
To estimate the upperbound of $R_{\rm SppS}$ conservatively,
we take the cross section at the Tevatron to be 2 pb in Eq. (\ref{eq:ratiospps}). 
Then, $R_{\rm SppS}$ have to be lower than 40.

\section{Non-resonant production}\label{sec:nonresonant}
We classify the effective operators by $X$'s
spin and its coupling to the standard model particles.
Note that in the case $W$ boson's interaction is same as that of the standard model,
$X$ must couple to the left-handed quarks.
We summarize the operators we considered in Table \ref{table:operators}.
We assume that only one operator mainly contributes to
the production of $X$ and $W$ boson at the Tevatron.
As for the operators not listed there,
we give some comments in the following subsections.
If we require the cross section of $WX$ signal to be 4 pb at the Tevatron,
we can get the required coupling constant.
However, the coupling constant depends on the normalization, for example,
definition of tensor which suppress $SU(3)_C\times SU(2)_L\times U(1)_Y$ indexes.
Therefore, in this section, we discuss the required decay width instead of the coupling constant.
We discuss physical quantity, then, we can avoid the confusion of the normalization.

Note that some operators do not make $X$ decay into dijet,
i.e., $G_{\m\n} W^{\m\n} \phi$, $B_{\m\n}W^{\m\n} \phi$,
$G_{\m\n} W^\m V^\n$ and $W_{\m\n}^3 W^\m V^\n$ in Table \ref{table:operators}.
In these cases, in addition to the original interaction, we must add another
interaction which forces $X$ to decay into dijet.
To avoid the suppression of the cross section by branching ratio,
we have to assume the decay interaction as strong as the original interaction.
This makes the analysis complicated,
because the decay interaction also contributes to the production process.
However, the original interaction always produces $X$ and $W$ boson by the SM gauge boson s-channel process.
Then, the production cross section is dominated by the process with the initial parton to be valence quark.
In order to dominate the production cross section of $WX$ signal for the original operator,
we have to assume the decay interaction forces $X$ to decay into sea quarks.
For the sea quark the factor $R_{\rm LHC}$ is larger than valence quark,
the operator introduced to makes $X$ decay enhances $R_{\rm LHC}$.
Therefore, adding the interaction which forces X decay into dijet improve the exclusion with the LHC.
Therefore, in the case the LHC
can exclude the operator we are considering without adding another
operator which contributes to the decay into dijet, we do not further analyze
with another operator.

In this section and the next section, we estimate a cross section at the hadron colliders by convoluting a parton level cross section with parton distribution functions.
We have used CTEQ 6.1 PDF~\cite{Stump:2003yu}.

\begin{table}
\begin{center}
\caption{List of the operators we considered.
$\phi,~V,~\psi_L$ is the new particle with spin 0, 1, 1/2 respectively.
Options for $U(1)_Y$
 charges are determined by whether $q_R$ is up-type or down-type quark.}
\begin{tabular}{|l||c|c|c||c|}
\hline
Interaction & $SU(3)_C$ & $SU(2)_L$ & $U(1)_Y$&Section \\\hline
\hline
$\phi q_L^{\dag}q_R$& 1 or 8 & 2 &	$\pm1/2$ &\ref{sec: scalar-yukawa}\\\hline
$\phi q_L^{\dag}q_R^c$&	$\bar{3}$ or 6 &  1 or 3 & 1/3 &\ref{sec: scalar-yukawa}\\\hline
$G_{\mu\nu}^aW^{\mu\nu}\phi^a$ & 8 & 3 & 0 &\ref{sec: scalar-GW}\\\hline
$B_{\mu\nu}W^{\mu\nu}\phi$ & 1 & 3 & 0 &\ref{sec: scalar-BW}\\\hline
\hline
$V^\mu q_L^{\dag}\bar{\sigma}_{\mu}q_L,$ & 1 or 8 & 1 or 3 & 0
	     &\ref{sec: vector-current}\\\hline
$ V^\mu q_L^{\dag}\bar{\sigma}_{\mu}q_L^c$ & $\bar{3}$ or 6 & 2 & 5/6 or
	     -1/6 &\ref{sec: vector-current}\\\hline
$V_{\mu\nu}q_L^{c\dag}\sigma^{\mu\nu}q_{L}$ & 3 or $\bar{6}$ & 1 or 3 &
	     -1/3 &\ref{sec: vector-Pauli}\\\hline
$V_{\mu\nu}q^{\dag}_R\sigma^{\mu\nu}q_{L}$ & 1 or 8 & 2 & $\pm1/2$
	     &\ref{sec: vector-Pauli}\\\hline
$G_{\mu\nu}^aW^{\mu}V^{a\nu}$ (arise from
 $G_{\mu\nu}^aD^{\mu}V^{a\nu}H$) & 8 & 2 & 1/2 &\ref{sec: vector-GW}\\\hline
$W^3_{\mu\nu}W^{\mu}V^{\nu}$ (arise from $W^i_{\mu\nu}(D^{\mu}V^{\nu})^i$) & 1 & 3 & 0 &\ref{sec: vector-BW}\\\hline
\hline
$G_{\mu\nu}q_L^{c\dag}\sigma^{\mu\nu}\psi_L$ & $\bar{3}$ or 6 or 15 & 2 & -1/6 &\ref{sec: spinor-Pauli}\\\hline
\end{tabular}
\label{table:operators}
\end{center}
\end{table}

\subsection{$X=\phi$ : scalar particle}
\label{sec: scalar}
\subsubsection{Quark- Quark- $\phi$}
\label{sec: scalar-yukawa}
The possible operators are
\begin{eqnarray}
\phi q_L^{\dag}q_R~~ {\rm or}~~ \phi q_L^{\dag}q_R^c.
\end{eqnarray}
and their conjugates.
These interaction terms give $W\phi$ signal by t-channel quark exchange diagrams.
The cross section of $Wjj$ signal is proportional to $N_\phi \Gamma(\phi\to qq)$,
where $\G(\phi\to qq)$ is a decay width of $\phi$,
and $N_\phi$ is the dimension of $\phi$ as a representation of $SU(3)_C$.
The required $N_\phi \Gamma(\phi\to qq)$ are summarized in Table \ref{tb: scalar-yukawa-decay width}.
Also, $R_{\rm LHC}$ and $R_{\rm SppS}$ is shown in Table \ref{tb: scalar-yukawa-LHC7/Tevatron}
and \ref{tb: scalar-yukawa-SppS/Tevatron}, respectively.
As is mentioned in Subsection \ref{sec: SppS}, we require that
  the decay width of $\phi$ is lower than 15 GeV when applying the exclusion
  criteria based on the predicted production cross section of $\phi$ at
  Sp${\rm \bar{p}}$S.
$R_{\rm SppS}$ does not depend on $N_\phi$,
but smaller $N_\phi$ can make the constraint loose because of large decay width.
In Table \ref{tb: scalar-yukawa-SppS/Tevatron},
some particle can not be excluded for smaller $SU(3)_C$ representation,
because the required decay width for the smaller representation is larger than 15 GeV.

All the operators are excluded from the ATLAS result unless the operators
are composed of only valence quarks and $\phi$.  Even for operators
composed of valence quarks, some of them are excluded by the UA2 result.

\subsubsection{Gluon- $W$ boson - $\phi$}
\label{sec: scalar-GW}
\begin{figure}[htbp]
\begin{center}
\includegraphics[scale=0.5]{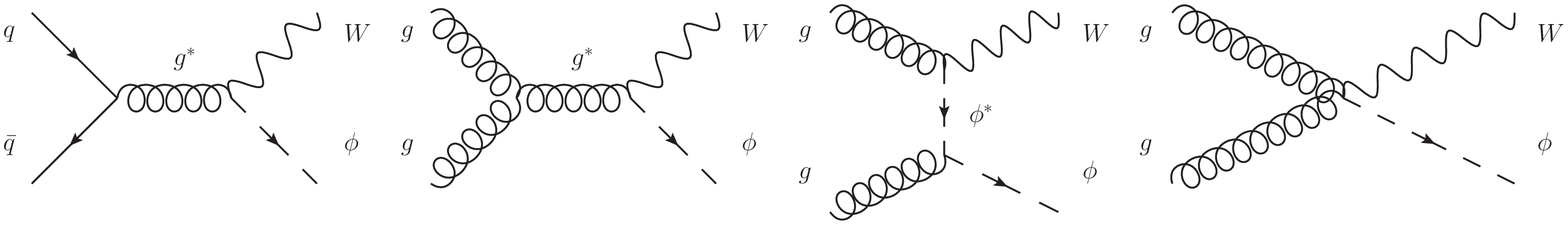}
\end{center}
\caption{The diagrams contributing to the production of $W\phi$ by the
 operator $G_{\mu\nu}^aW^{\mu\nu}\phi^a$ at hadron colliders.}
\label{fig: GWphi-fig}
\end{figure}

The possible operators are
\begin{eqnarray}
 G_{\mu\nu}^aW^{\mu\nu}\phi^a
\end{eqnarray}
and its conjugate. These interaction terms give $W\phi$ signal by the
diagrams shown in Fig.~\ref{fig: GWphi-fig}. 
The required coupling for 4 pb at the Tevatron can be translated into the decay width
$\Gamma(\phi\rightarrow gW)$ of $8.5\times 10^{-2}$ GeV. This corresponds to the cutoff
$\Lambda$ of 850 GeV with the normalization $\frac{1}{\Lambda}
G_{\mu\nu}^aW^{\mu\nu}\phi^a$.
We estimate $R_{\rm LHC}$ is $8.6$, then, this is excluded by the ATLAS result at 95 \%
C.L.~\cite{ATLAS_wjj}.  As we mentioned at the beginning of this section,
we do not analyze further with another operator which make $\phi$ decay
into dijet.

\subsubsection{Photon or Z boson - $W$ boson - $\phi$}
\label{sec: scalar-BW}
The possible operators are 
\begin{eqnarray}
 B_{\mu\nu}W^{\mu\nu}\phi=(-{\rm sin}\theta_w Z_{\mu\nu}+{\rm cos}\theta_wF_{\mu\nu})W^{\mu\nu}\phi
\end{eqnarray}
and its conjugate. These interaction terms give $W\phi$ signal by
s-channel photon and $Z$ boson exchange. The required decay width
$\Gamma(\phi\rightarrow\gamma W)$ for 4 pb at the Tevatron is $7.3$ GeV.
This corresponds to the cutoff
$\Lambda$ of 100 GeV with the normalization $\frac{1}{\Lambda}B_{\mu\nu}W^{\mu\nu}\phi$.
We estimate $R_{\rm LHC}$ is $3.4$.
Though this is not excluded by the ATLAS result,
cut off of 100 GeV is not acceptable as a higher dimensional operator.
Adding another operator to make $\phi$ decay into dijet and
recalculating the required decay width inevitably lower
this cut off, we do not further analyze.
We can think of an operator like $W_{\mu\nu}W^{\mu\nu}\phi$.  However,
such operator will result in an unacceptably lower cutoff, too.

\subsection{$X=V$ : vector particle}
Since there are a lot of possible operators to explain the $WV$
production at the CDF, we pick up representative operators.  Some variation
of them are discussed in the text.
\subsubsection{Quark - Quark - $V$ : $V^{\mu}J_{\mu}$ type}
\label{sec: vector-current}
The possible operators are
\begin{eqnarray}
V^\mu q_L^{\dag}\bar{\sigma}_{\mu}q_L~~{\rm or}~~  V^\mu q_L^{\dag}\bar{\sigma}_{\mu}q_L^c
\end{eqnarray}
and their conjugates.
These interaction terms give $W\phi$ signal by t-channel quark exchange diagrams.
The following analysis is same as that
of Subsection \ref{sec: scalar}.
The required $N_V \Gamma(V\to qq)$ are summarized in Table \ref{tb: vector-current-decay width}.
$R_{\rm LHC}$ and $R_{\rm SppS}$ are shown
in Table \ref{tb: vector-current-LHC7/Tevatron} and \ref{tb: vector-current-SppS/Tevatron}, respectively.
As is mentioned in Subsubsection \ref{sec: scalar-yukawa},
the exclusion based of the UA2 result depend on which $SU(3)_C$ representation $V$ obeys.
All the operators which do not contain valence quarks are excluded by the ATLAS result.

\subsubsection{Quark - quark- $V$: Pauli term type}
\label{sec: vector-Pauli}
The possible operators are
\begin{eqnarray}
 V_{\mu\nu}q_R^{c\dag}\sigma^{\mu\nu}q_{L}~~{\rm or}~~ V_{\mu\nu}q^{\dag}_R\sigma^{\mu\nu}q_{L}
\end{eqnarray}
and their conjugates where
$\sigma^{\mu\nu}=\frac{i}{2}(\sigma^{\mu}\bar{\sigma}^{\nu}-\sigma^{\nu}\bar{\sigma}^{\mu})$.
These interaction terms give $W\phi$ signal by t-channel quark exchange diagrams.
The following analysis is same as that of \ref{sec: scalar}.
The required $N_V \Gamma(V\to qq)$ are summarized in Table \ref{tb: vector-Pauli-decay width}.

This can be also expressed in terms of the strength of the coupling.
We have shown the required cut
off $\Lambda$ in Table \ref{tb: vector-Pauli-cutoff} with the normalization 
\begin{eqnarray}
 \frac{\sqrt{3}}{\sqrt{N_V}\Lambda}C_{ij}^rV_{\mu\nu}^rq_i^{\dag}\sigma^{\mu\nu}q_j,
\end{eqnarray}  
where $q$ is two component quark field, $i$,$j$ are the indexes of the color of quarks, $r$
is the index of color of $V$, $C_{ij}^r$ is the Clebsch-Gordan
coefficients of the expansion $3(\bar{3})\otimes3(\bar{3})$ to the
representation of $SU(3)_c$ which $V$ obeys. Here, the Clebsch-Gordan
coefficients are normalized to satisfy
$\sum_{i,j}C_{ij}^rC_{ij}^{r'*}=\delta^{rr'}$ and $\sum_{r}C_{ij}^rC_{i'j'}^{r*}=\delta_{ii'}\delta_{jj'}$.
$R_{\rm LHC}$ and $R_{\rm SppS}$ are shown
in Table \ref{tb: vector-Pauli-LHC7/Tevatron} and \ref{tb: vector-Pauli-SppS/Tevatron}, respectively.
As is mentioned in Subsubsection \ref{sec: scalar-yukawa},
the exclusion based of the UA2 result depend on which $SU(3)_c$ representation $V$ obeys.
All the operators which do not contain valence quarks are excluded by
the ATLAS result.

\subsubsection{Gluon - $W$ boson - $V$}
\label{sec: vector-GW}
\begin{figure}
\begin{center}
\includegraphics[scale=0.5]{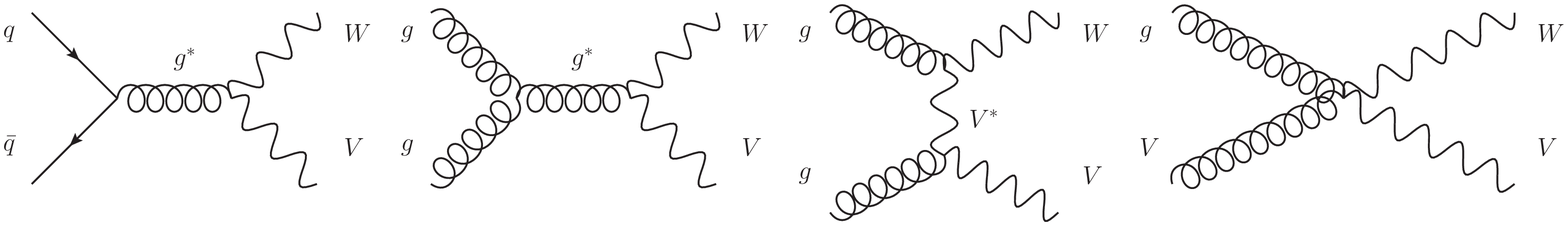}
\end{center}
\caption{The diagrams contributing to the production of $WV$ by the
 operator $G_{\mu\nu}^aW^{\mu}V^{a\nu}$ at hadron colliders.}
\label{fig: GWV-fig}
\end{figure}

The possible operators are
\begin{eqnarray}
 G_{\mu\nu}^aW^{\mu}V^{a\nu}
\end{eqnarray}
and its conjugate. These interaction terms give $W\phi$ signal by the
diagrams shown in Fig.~\ref{fig: GWV-fig}. The required decay width
$\Gamma(V\rightarrow gW)$ is $3.4\times10^{-2}$ GeV.
We estimate $R_{\rm LHC}$ is $1.4 \times 10^1$, then, this is excluded by the ATLAS result at 95
\% C.L.~\cite{ATLAS_wjj}. As we mentioned at the beginning of this section,
we do not analyze further with another operator which make $\phi$ decay
into dijet since.

\subsubsection{Photon or Z boson - $W$ boson - $V$}
\label{sec: vector-BW}
The possible operators are
\begin{eqnarray}
 W^3_{\mu\nu}W^{\mu}V^{\nu}=({\rm cos}\theta_wZ_{\mu\nu}+{\rm sin}\theta_wF_{\mu\nu})W^{\mu}V^{\nu}
\end{eqnarray}
and its conjugate. These interaction terms give $WV$ signal by s-channel
photon and $Z$ boson exchange. The required decay width
$\Gamma(V\rightarrow \gamma W)$ is $1.1$ GeV.
$R_{\rm LHC}$ is $6.8$.  This value is allowed by the ATLAS result at 95
\% C.L.~\cite{ATLAS_wjj}.

As we mentioned at the beginning of this
section, we must add another operator to make $V$ decay into dijet.
This can be safely introduced without contradiction to Sp${\rm\bar{p}}$S
nor LHC experiment.  For example, take the decay width to $c\bar{s}$
to be 2 GeV and to $\gamma W$ to be 2 GeV so that production cross
section of $V$ and $W$ boson times the branching fraction of $V$ to
dijet is 4 pb at the Tevatron. In this case,
We estimate $R_{\rm SppS}$ is 0.1, which is not excluded.
We estimate $R_{\rm LHC}$ is 7, which is also not excluded.

We can think of operators like $W_{\mu\nu}W_{\mu}^3V_{\nu}$, $B_{\mu\nu}W^{\mu}V^{\nu}$.  These
operators result in the similar cross section ratio. Therefore, such
operators are still allowed, too.

\subsection{$X=\psi$ : spinor particle}

\subsubsection{Gluon - quark - $\psi$}
\label{sec: spinor-Pauli}
The possible operators are
\begin{eqnarray}
G_{\mu\nu}q_R^{c\dag}\sigma^{\mu\nu}\psi_L\
\end{eqnarray}
and their conjugates.
These interaction terms give $W\phi$ signal by t-channel quark exchange diagrams.
The following analysis is same as that of Subsection \ref{sec: scalar}.
The required $N_\psi \Gamma(\psi\to qg)$ are summarized in Table \ref{tb: spinor-Pauli-decay width}.
The corresponding cut off $\Lambda$ is shown in Table \ref{tb: spinor-Pauli-cutoff} in GeV with the normalization
\begin{eqnarray}
 \frac{2}{\sqrt{N_{\psi}}\Lambda}C_{ai}^rG_{\mu\nu}^aq^{\dag}_i\sigma^{\mu\nu}\psi_L^r,
\end{eqnarray}
where $q$ is the two component quark field, $i$ is the index of the color of quark, $r$ is
the index of the color of $\psi$ and $C_{ai}^r$ is the Clebsch-Gordan
coefficients of the expansion $8\otimes 3(\bar{3})$ to the
representation of $SU(3)_c$ which $\psi$ obeys. Here, the Clebsch-Gordan
coefficients are normalized to satisfy
$\sum_{a,i}C_{ai}^rC_{ai}^{r'*}=\delta^{rr'}$ and
$\sum_{r}C_{ai}^rC_{a'i'}^{r*}=\delta_{aa'}\delta_{ii'}$.
$R_{\rm LHC}$ and $R_{\rm SppS}$ are shown
in Table \ref{tb: spinor-Pauli-LHC7/Tevatron} and \ref{tb: spinor-Pauli-SppS/Tevatron}, respectively.
All the operators are excluded by the ATLAS result at 95 \% C.L.~\cite{ATLAS_wjj}. 
As is mentioned in Subsubsection \ref{sec: scalar-yukawa}, the exclusion based of the UA2 result depend on which
$SU(3)_c$ representation $V$ obeys.

We can think of operators like $W^{\mu}q^{\dag}\sigma_{\mu}\psi$.
However, since gluon in necessary for initial state in order to produce
$W\psi$, such operator will result in exclusion by the ATLAS result, too.

\section{Resonant production}\label{sec:resonant}
\label{sec: resonance}

In this section, we consider the case in which $WX$ are produced by the decay of another particle $Y$, such as:
\begin{eqnarray}
p+\bar p (p) ~\to~ Y ~\to~ W+X.
\end{eqnarray}
In calculating the production cross section, we use the narrow width approximation.\footnote{
For some combinations of assumed initial partons, the required decay width for 4 pb at the Tevatron might go be beyond the validity of the narrow
width approximation.  In such a case, as is shown later, the ratio of the production cross section of $Y$ at the LHC with $R_{\rm LHC}$ is greater than 40.
Since such operators are excluded by ATLAS
result, we do not have to consider such operators
from the beginning.}
Then, the production cross section is proportional to the decay width of $Y\to ij$.
Here, $i$ and $j$ are initial partons.
The production cross section is given by,
\begin{eqnarray}
\sigma(Y) = C(i,j) \times N_Y \Gamma(Y\to ij),
\end{eqnarray}
where $N_Y$ is the degree of freedom of the particle $Y$.
The coefficient $C(i,j)$ is determined by the parton distribution
function of incoming hadrons and do not depend on the
detailed property of $Y$ such as the spin and quantum numbers.
Therefore, for each combination of $i$ and $j$,
we can calculate the required decay width for 4 pb at the Tevatron.
The required $N_Y\Gamma(Y\rightarrow ij)$ are summarized in
Table~\ref{tb: resonance-decaywidth}.

$R_{\rm LHC}$, which depend only on the initial partons, is shown
in Table \ref{tb: resonance-LHC7/Tevatron}.
The operators including only valence quark are not excluded by the ATLAS result.

\clearpage

\begin{table}[h]
\begin{center}
\caption{
($\phi q_L^iq^j$, \ref{sec: scalar-yukawa})
The required $N_\phi \Gamma(\phi\to qq)$ / 1 GeV for the $W\phi$ cross section to be 4 pb at the Tevatron.
 }
\begin{tabular}[tb]{|c||c|c|c|c|c|}
\hline
\backslashbox{$i$}{$j$}  &  $\bar{b}$  &  $\bar{c}$  &  $\bar{s}$  & $\bar{u}$  &  $\bar{d}$  \\\hline\hline
$\bar{b}$& $5.3\times 10^{5}$ & $9.7\times 10^{2}$ & $9.1\times 10^{2}$ & $1.9\times 10^{2}$ & $8.3\times 10^{1}$ \\\hline
$\bar{c}$& $9.7\times 10^{2}$ & $2.8\times 10^{2}$ & $2.3\times 10^{2}$ & $2.1\times 10^{1}$ & $2.7\times 10^{1}$ \\\hline
$\bar{s}$& $9.1\times 10^{2}$ & $2.3\times 10^{2}$ & $1.6\times 10^{2}$ & $2.6\times 10^{1}$ & $2.2\times 10^{1}$ \\\hline
$\bar{u}$& $1.9\times 10^{2}$ & $2.1\times 10^{1}$ & $2.6\times 10^{1}$ & $6.8$& $9.9$ \\\hline
$\bar{d}$& $8.3\times 10^{1}$ & $2.7\times 10^{1}$ & $2.2\times 10^{1}$ & $9.9$& $6.9$ \\\hline\hline
$d$& $8.3\times 10^{1}$ & $5.0\times 10^{1}$ & $3.1\times 10^{1}$ & $1.8$& $3.8$ \\\hline
$u$& $1.9\times 10^{2}$ & $1.2\times 10^{2}$ & $7.0\times 10^{1}$ & $3.8$& $8.4$ \\\hline
$s$& $9.1\times 10^{2}$ & $5.4\times 10^{2}$ & $3.3\times 10^{2}$ & $2.0\times 10^{1}$ & $4.4\times 10^{1}$ \\\hline
$c$& $9.7\times 10^{2}$ & $5.6\times 10^{2}$ & $3.4\times 10^{2}$ & $2.1\times 10^{1}$ & $4.6\times 10^{1}$ \\\hline
$b$& $1.1\times 10^{6}$ & $6.2\times 10^{5}$ & $3.7\times 10^{5}$ & $2.3\times 10^{4}$ & $5.2\times 10^{4}$ \\\hline
 \end{tabular}
\label{tb: scalar-yukawa-decay width}
\vspace{1cm}
\caption{
($\phi q_L^iq^j$, \ref{sec: scalar-yukawa})
The enhancement factor $R_{\rm LHC}$ defined in Eq. (\ref{eq:ratio}).
The elements marked $\circ$ and written in blue letters are NOT excluded by
 the ATLAS result.}
\begin{tabular}[tb]{|c||c|c|c|c|c|}
\hline
\backslashbox{$i$}{$j$}  &  $\bar{b}$  &  $\bar{c}$  &  $\bar{s}$  & $\bar{u}$  &  $\bar{d}$ \\\hline\hline
$\bar{b}$& $6.5\times 10^{1}$ & $5.9\times 10^{1}$ & $4.8\times 10^{1}$ & $3.8\times 10^{1}$ & $2.9\times 10^{1}$ \\\hline
$\bar{c}$& $5.9\times 10^{1}$ & $5.3\times 10^{1}$ & $4.5\times 10^{1}$ & $2.4\times 10^{1}$ & $2.7\times 10^{1}$ \\\hline
$\bar{s}$& $4.8\times 10^{1}$ & $4.5\times 10^{1}$ & $3.8\times 10^{1}$ & $2.7\times 10^{1}$ & $2.4\times 10^{1}$ \\\hline
$\bar{u}$& $3.8\times 10^{1}$ & $2.4\times 10^{1}$ & $2.7\times 10^{1}$ & $3.0\times 10^{1}$ & $2.4\times 10^{1}$ \\\hline
$\bar{d}$& $2.9\times 10^{1}$ & $2.7\times 10^{1}$ & $2.4\times 10^{1}$ & $2.4\times 10^{1}$ & $3.0\times 10^{1}$ \\\hline\hline
$d$& $2.9\times 10^{1}$ & $2.6\times 10^{1}$ & $2.1\times 10^{1}$ &
		 \textcolor{blue}{$\circ~2.9$} & \textcolor{blue}{$\circ~5.0$}\\\hline
$u$& $3.8\times 10^{1}$ & $3.4\times 10^{1}$ & $2.8\times 10^{1}$ &
		 \textcolor{blue}{$\circ~5.1$} & \textcolor{blue}{$\circ~8.0$} \\\hline
$s$& $4.8\times 10^{1}$ & $4.3\times 10^{1}$ & $3.8\times 10^{1}$ & $1.2\times 10^{1}$ & $1.6\times 10^{1}$ \\\hline
$c$& $5.9\times 10^{1}$ & $5.3\times 10^{1}$ & $4.7\times 10^{1}$ & $1.6\times 10^{1}$ & $2.2\times 10^{1}$ \\\hline
$b$& $6.5\times 10^{1}$ & $5.9\times 10^{1}$ & $5.3\times 10^{1}$ & $2.1\times 10^{1}$ & $2.7\times 10^{1}$ \\\hline
 \end{tabular}
\label{tb: scalar-yukawa-LHC7/Tevatron}
\end{center}
\end{table}
\begin{table}[h]
\begin{center}
\caption{($\phi q_L^iq^j$, \ref{sec: scalar-yukawa})
The enhancement factor $R_{\rm SppS}$ defined in Eq. (\ref{eq:ratiospps}).
The elements marked $\bullet$ and written in red letters are
excluded by the UA2 result at 90 \% C.L.~\cite{Alitti:1993pn} for $\phi$ which
  obeys any representation of $SU(3)_C$.
$\ast$ and green letters are excluded only $SU(3)_C$ higher representation,
i.e., $8$ or $6(\bar{6})$.
}
\begin{tabular}[tb]{|c||c|c|c|c|c|}
\hline
\backslashbox{$i$}{$j$}  &  $\bar{b}$  &  $\bar{c}$  &  $\bar{s}$  & $\bar{u}$  &  $\bar{d}$ \\\hline\hline
$\bar{b}$& $7.4\times 10^{3}$ & $1.3\times 10^{1}$ & $2.0\times 10^{1}$ & $1.1\times 10^{2}$ & $1.8\times 10^{1}$ \\\hline
$\bar{c}$& $1.3\times 10^{1}$ & $1.5\times 10^{1}$ & $10$& $2.3\times 10^{1}$ & $1.1\times 10^{1}$ \\\hline
$\bar{s}$& $2.0\times 10^{1}$ & $10$& $2.4\times 10^{1}$ & \textcolor{green}{$\ast~4.9\times 10^{1}$} & $1.6\times 10^{1}$ \\\hline
$\bar{u}$& $1.1\times 10^{2}$ & $2.3\times 10^{1}$ & \textcolor{green}{$\ast~4.9\times 10^{1}$} & \textcolor{red}{$\bullet~7.3\times 10^{1}$} & \textcolor{red}{$\bullet~4.3\times 10^{1}$} \\\hline
$\bar{d}$& $1.8\times 10^{1}$ & $1.1\times 10^{1}$ & $1.6\times 10^{1}$ & \textcolor{red}{$\bullet~4.3\times 10^{1}$} & $3.2\times 10^{1}$ \\\hline\hline
$d$& $1.8\times 10^{1}$ & $2.1\times 10^{1}$ & $2.2\times 10^{1}$ & \textcolor{red}{$\bullet~4.2\times 10^{1}$} & $3.6\times 10^{1}$ \\\hline
$u$& $1.2\times 10^{2}$ & $1.3\times 10^{2}$ & $1.3\times 10^{2}$ & \textcolor{red}{$\bullet~2.1\times 10^{2}$} & \textcolor{green}{$\ast~2.0\times 10^{2}$} \\\hline
$s$& $2.0\times 10^{1}$ & $2.3\times 10^{1}$ & $2.4\times 10^{1}$ & $3.8\times 10^{1}$ & $3.1\times 10^{1}$ \\\hline
$c$& $1.3\times 10^{1}$ & $1.5\times 10^{1}$ & $1.5\times 10^{1}$ & $2.2\times 10^{1}$ & $1.9\times 10^{1}$ \\\hline
$b$& $7.4\times 10^{3}$ & $8.4\times 10^{3}$ & $8.4\times 10^{3}$ & $1.4\times 10^{4}$ & $1.1\times 10^{4}$\\\hline
 \end{tabular}
\label{tb: scalar-yukawa-SppS/Tevatron}
\vspace{1cm}
\caption{
($V_{\mu}q_L^i\sigma^{\mu}q^j$, \ref{sec: vector-current})
The required $N_V \Gamma(V\to qq)$ / 1 GeV for 4 pb at the Tevatron.
 }
\begin{tabular}[tb]{|c||c|c|c|c|c|}
\hline
\backslashbox{$i$}{$j$}  &  $\bar{b}$  &  $\bar{c}$  &  $\bar{s}$  & $\bar{u}$  &  $\bar{d}$ \\\hline\hline
$\bar{b}$& $2.5\times 10^{5}$ & $4.5\times 10^{2}$ & $4.2\times 10^{2}$ & $8.9\times 10^{1}$ & $3.8\times 10^{1}$ \\\hline
$\bar{c}$& $4.5\times 10^{2}$ & $1.3\times 10^{2}$ & $9.1\times 10^{1}$ & $9.7$& $1.2\times 10^{1}$ \\\hline
$\bar{s}$& $4.2\times 10^{2}$ & $9.1\times 10^{1}$ & $7.6\times 10^{1}$ & $1.1\times 10^{1}$ & $1.0\times 10^{1}$ \\\hline
$\bar{u}$& $8.9\times 10^{1}$ & $9.7$& $1.1\times 10^{1}$ & $3.1$& $3.5$ \\\hline
$\bar{d}$& $3.8\times 10^{1}$ & $1.2\times 10^{1}$ & $1.0\times 10^{1}$ & $3.5$& $3.2$ \\\hline\hline
$d$& $3.8\times 10^{1}$ & $2.3\times 10^{1}$ & $1.4\times 10^{1}$ & $7.6\times 10^{-1}$ & $1.7$ \\\hline
$u$& $9.0\times 10^{1}$ & $5.3\times 10^{1}$ & $3.2\times 10^{1}$ & $1.7$& $3.7$\\\hline
$s$& $4.2\times 10^{2}$ & $2.5\times 10^{2}$ & $1.5\times 10^{2}$ & $8.6$& $1.9\times 10^{1}$ \\\hline
$c$& $4.5\times 10^{2}$ & $2.6\times 10^{2}$ & $1.6\times 10^{2}$ & $9.2$& $2.1\times 10^{1}$ \\\hline
$b$& $5.0\times 10^{5}$ & $2.9\times 10^{5}$ & $1.7\times 10^{5}$ & $1.0\times 10^{4}$ & $2.4\times 10^{4}$ \\\hline
 \end{tabular}
\label{tb: vector-current-decay width}
\end{center}
\end{table}
\begin{table}[h]
\begin{center}
\caption{($V_{\mu}q_L^i\sigma^{\mu}q^j$, \ref{sec: vector-current})
The enhancement factor $R_{\rm LHC}$ for $V_{\mu}q_L^i\sigma^{\mu}q^j$.
}
\begin{tabular}[tb]{|c||c|c|c|c|c|}
\hline
\backslashbox{$i$}{$j$}  &  $\bar{b}$  &  $\bar{c}$  &  $\bar{s}$  & $\bar{u}$  &  $\bar{d}$ \\\hline\hline
$\bar{b}$& $9.1\times 10^{1}$ & $8.3\times 10^{1}$ & $6.7\times 10^{1}$ & $5.3\times 10^{1}$ & $4.2\times 10^{1}$ \\\hline
$\bar{c}$& $8.3\times 10^{1}$ & $7.5\times 10^{1}$ & $5.4\times 10^{1}$ & $3.6\times 10^{1}$ & $3.8\times 10^{1}$\\\hline
$\bar{s}$& $6.7\times 10^{1}$ & $5.4\times 10^{1}$ & $5.4\times 10^{1}$ & $3.8\times 10^{1}$ & $3.6\times 10^{1}$ \\\hline
$\bar{u}$& $5.3\times 10^{1}$ & $3.6\times 10^{1}$ & $3.8\times 10^{1}$ & $5.3\times 10^{1}$ & $3.0\times 10^{1}$ \\\hline
$\bar{d}$& $4.2\times 10^{1}$ & $3.8\times 10^{1}$ & $3.6\times 10^{1}$ & $3.0\times 10^{1}$ & $5.4\times 10^{1}$ \\\hline\hline
$d$& $4.2\times 10^{1}$ & $3.7\times 10^{1}$ & $3.1\times 10^{1}$ & \textcolor{blue}{$\circ~4.0$}& \textcolor{blue}{$\circ~7.1$} \\\hline
$u$& $5.4\times 10^{1}$ & $4.8\times 10^{1}$ & $4.1\times 10^{1}$ & \textcolor{blue}{$\circ~7.2$}& $1.1\times 10^{1}$ \\\hline
$s$& $6.7\times 10^{1}$ & $6.1\times 10^{1}$ & $5.4\times 10^{1}$ & $1.7\times 10^{1}$ & $2.3\times 10^{1}$ \\\hline
$c$& $8.3\times 10^{1}$ & $7.5\times 10^{1}$ & $6.6\times 10^{1}$ & $2.4\times 10^{1}$ & $3.1\times 10^{1}$ \\\hline
$b$& $9.1\times 10^{1}$ & $8.3\times 10^{1}$ & $7.5\times 10^{1}$ & $3.0\times 10^{1}$ & $3.9\times 10^{1}$ \\\hline
 \end{tabular}
\label{tb: vector-current-LHC7/Tevatron}
\vspace{1cm}
\caption{($V_{\mu}q_L^i\sigma^{\mu}q^j$, \ref{sec: vector-current})
The enhancement factor $R_{\rm SppS}$ for $V_{\mu}q_L^i\sigma^{\mu}q^j$.
}
\begin{tabular}[tb]{|c||c|c|c|c|c|}
\hline
\backslashbox{$i$}{$j$}  &  $\bar{b}$  &  $\bar{c}$  &  $\bar{s}$  & $\bar{u}$  &  $\bar{d}$ \\\hline\hline
$\bar{b}$& $5.2\times 10^{3}$ & $5.4\times 10^{3}$ & $5.9\times 10^{3}$ & $1.1\times 10^{4}$ & $9.4\times 10^{3}$ \\\hline
$\bar{c}$& $9.2$& $1.0\times 10^{1}$ & $9.6$& $1.9\times 10^{1}$ & $1.6\times 10^{1}$ \\\hline
$\bar{s}$& $1.4\times 10^{1}$ & $1.6\times 10^{1}$ & $1.6\times 10^{1}$ & \textcolor{green}{$\ast~5.1\times 10^{1}$} & $3.8\times 10^{1}$ \\\hline
$\bar{u}$& $8.0\times 10^{1}$ & $8.7\times 10^{1}$ & \textcolor{green}{$\ast~9.2\times 10^{1}$} & \textcolor{red}{$\bullet~5.0\times 10^{1}$} & \textcolor{red}{$\bullet~6.6\times 10^{1}$} \\\hline
$\bar{d}$& $1.3\times 10^{1}$ & $1.4\times 10^{1}$ & $1.5\times 10^{1}$ & $3.5\times 10^{1}$ & $2.2\times 10^{1}$ \\\hline\hline
$d$& $1.3\times 10^{1}$ & $6.8$& $8.7$& $2.2\times 10^{1}$ & $1.2\times 10^{1}$ \\\hline
$u$& $7.9\times 10^{1}$ & $1.3\times 10^{1}$ & $1.9\times 10^{1}$ & \textcolor{red}{$\bullet~6.8\times 10^{1}$} & $2.2\times 10^{1}$ \\\hline
$s$& $1.4\times 10^{1}$ & $6.3$& $8.2$& $1.9\times 10^{1}$ & $8.7$ \\\hline
$c$& $9.2$& $5.1$& $6.3$& $1.3\times 10^{1}$ & $6.8$\\\hline
$b$& $2.6\times 10^{3}$ & $9.2$& $1.4\times 10^{1}$ & $7.9\times 10^{1}$ & $1.3\times 10^{1}$ \\\hline
 \end{tabular}
\label{tb: vector-current-SppS/Tevatron}
\end{center}
\end{table}
\begin{table}[h]
\begin{center}
\caption{($V_{\mu\nu}q_L^i\sigma^{\mu\nu}q^j$, \ref{sec: vector-Pauli})
The required $N_V \Gamma(V\to qq)$ / 1 GeV for 4 pb at the Tevatron.
 }
\begin{tabular}[tb]{|c||c|c|c|c|c|}
\hline
\backslashbox{$i$}{$j$}  &  $\bar{b}$  &  $\bar{c}$  &  $\bar{s}$  & $\bar{u}$  &  $\bar{d}$ \\\hline\hline
$\bar{b}$& $6.8\times 10^{4}$ & $1.2\times 10^{2}$ & $1.1\times 10^{2}$ & $2.4\times 10^{1}$ & $1.0\times 10^{1}$ \\\hline
$\bar{c}$& $1.2\times 10^{2}$ & $3.6\times 10^{1}$ & $3.2\times 10^{1}$ & $2.6$& $3.3$\\\hline
$\bar{s}$& $1.1\times 10^{2}$ & $3.2\times 10^{1}$ & $2.0\times 10^{1}$ & $3.2$& $2.7$\\\hline
$\bar{u}$& $2.4\times 10^{1}$ & $2.6$& $3.2$& $8.2\times 10^{-1}$ & $1.4$\\\hline
$\bar{d}$& $1.0\times 10^{1}$ & $3.3$& $2.7$& $1.4$& $8.3\times 10^{-1}$ \\\hline\hline
$d$& $1.0\times 10^{1}$ & $6.0$& $3.7$& $1.8\times 10^{-1}$ & $4.2\times 10^{-1}$ \\\hline
$u$& $2.4\times 10^{1}$ & $1.4\times 10^{1}$ & $8.7$& $4.1\times 10^{-1}$ & $9.5\times 10^{-1}$ \\\hline
$s$& $1.1\times 10^{2}$ & $6.7\times 10^{1}$ & $4.1\times 10^{1}$ & $2.2$& $5.0$ \\\hline
$c$& $1.2\times 10^{2}$ & $7.2\times 10^{1}$ & $4.3\times 10^{1}$ & $2.4$& $5.6$ \\\hline
$b$& $1.4\times 10^{5}$ & $7.8\times 10^{4}$ & $4.8\times 10^{4}$ & $2.7\times 10^{3}$ & $6.3\times 10^{3}$ \\\hline
 \end{tabular}
\label{tb: vector-Pauli-decay width}
\vspace{1cm}
\caption{($V_{\mu\nu}q_L^i\sigma^{\mu\nu}q^j$, \ref{sec: vector-Pauli})
The required cut off $\Lambda$ in GeV for 4 pb at the Tevatron.
See the text for the definition of $\Lambda$.
 }
\begin{tabular}[tb]{|c||c|c|c|c|c|}
\hline
\backslashbox{$i$}{$j$}  &  $\bar{b}$  &  $\bar{c}$  &  $\bar{s}$  & $\bar{u}$  &  $\bar{d}$ \\\hline\hline
$\bar{b}$& $1.4$& $4.7\times 10^{1}$ & $4.9\times 10^{1}$ & $1.1\times 10^{2}$ & $1.6\times 10^{2}$ \\\hline
$\bar{c}$& $4.7\times 10^{1}$ & $6.1\times 10^{1}$ & $9.2\times 10^{1}$ & $3.2\times 10^{2}$ & $2.8\times 10^{2}$ \\\hline
$\bar{s}$& $4.9\times 10^{1}$ & $9.2\times 10^{1}$ & $8.1\times 10^{1}$ & $2.9\times 10^{2}$ & $3.2\times 10^{2}$ \\\hline
$\bar{u}$& $1.1\times 10^{2}$ & $3.2\times 10^{2}$ & $2.9\times 10^{2}$ & $4.0\times 10^{2}$ & $4.4\times 10^{2}$ \\\hline
$\bar{d}$& $1.6\times 10^{2}$ & $2.8\times 10^{2}$ & $3.2\times 10^{2}$ & $4.4\times 10^{2}$ & $4.0\times 10^{2}$ \\\hline\hline
$d$& $1.6\times 10^{2}$ & $2.1\times 10^{2}$ & $2.7\times 10^{2}$ & $1.2\times 10^{3}$ & $8.0\times 10^{2}$ \\\hline
$u$& $1.1\times 10^{2}$ & $1.4\times 10^{2}$ & $1.8\times 10^{2}$ & $8.1\times 10^{2}$ & $5.3\times 10^{2}$ \\\hline
$s$& $4.9\times 10^{1}$ & $6.4\times 10^{1}$ & $8.1\times 10^{1}$ & $3.5\times 10^{2}$ & $2.3\times 10^{2}$ \\\hline
$c$& $4.7\times 10^{1}$ & $6.1\times 10^{1}$ & $7.9\times 10^{1}$ & $3.4\times 10^{2}$ & $2.2\times 10^{2}$ \\\hline
$b$& $1.4$& $1.9$& $2.4$& $10$& $6.5$\\\hline
 \end{tabular}
\label{tb: vector-Pauli-cutoff}
\end{center}
\end{table}
\begin{table}[h]
\begin{center}
\caption{($V_{\mu\nu}q_L^i\sigma^{\mu\nu}q^j$, \ref{sec: vector-Pauli})
The enhancement factor $R_{\rm LHC}$ for $V_{\mu\nu}q_L^i\sigma^{\mu\nu}q^j$.
}
\begin{tabular}[tb]{|c||c|c|c|c|c|}
\hline
\backslashbox{$i$}{$j$}  &  $\bar{b}$  &  $\bar{c}$  &  $\bar{s}$  & $\bar{u}$  &  $\bar{d}$ \\\hline\hline
 $\bar{b}$& $9.5\times 10^{1}$ & $8.8\times 10^{1}$ & $7.2\times 10^{1}$ & $6.0\times 10^{1}$ & $4.8\times 10^{1}$ \\\hline
 $\bar{c}$& $8.8\times 10^{1}$ & $8.0\times 10^{1}$ & $5.1\times 10^{1}$ & $4.3\times 10^{1}$ & $4.4\times 10^{1}$ \\\hline
 $\bar{s}$& $7.2\times 10^{1}$ & $5.1\times 10^{1}$ & $5.9\times 10^{1}$ & $4.4\times 10^{1}$ & $4.4\times 10^{1}$ \\\hline
 $\bar{u}$& $6.0\times 10^{1}$ & $4.3\times 10^{1}$ & $4.4\times 10^{1}$ & $7.5\times 10^{1}$ & $3.8\times 10^{1}$ \\\hline
 $\bar{d}$& $4.8\times 10^{1}$ & $4.4\times 10^{1}$ & $4.4\times 10^{1}$ & $3.8\times 10^{1}$ & $7.6\times 10^{1}$ \\\hline\hline
$d$& $4.8\times 10^{1}$ & $4.3\times 10^{1}$ & $3.7\times 10^{1}$ & \textcolor{blue}{$\circ~4.5$}& \textcolor{blue}{$\circ~8.2$} \\\hline
$u$& $6.0\times 10^{1}$ & $5.4\times 10^{1}$ & $4.7\times 10^{1}$ & \textcolor{blue}{$\circ~8.3$}& $1.3\times 10^{1}$ \\\hline
$s$& $7.2\times 10^{1}$ & $6.5\times 10^{1}$ & $5.9\times 10^{1}$ & $1.9\times 10^{1}$ & $2.5\times 10^{1}$ \\\hline
$c$& $8.8\times 10^{1}$ & $8.0\times 10^{1}$ & $7.2\times 10^{1}$ & $2.8\times 10^{1}$ & $3.6\times 10^{1}$ \\\hline
$b$& $9.5\times 10^{1}$ & $8.6\times 10^{1}$ & $7.9\times 10^{1}$ & $3.4\times 10^{1}$ & $4.3\times 10^{1}$ \\\hline
 \end{tabular}
\label{tb: vector-Pauli-LHC7/Tevatron}
\vspace{1cm}
\caption{($V_{\mu\nu}q_L^i\sigma^{\mu\nu}q^j$, \ref{sec: vector-Pauli})
The enhancement factor $R_{\rm SppS}$ for $V_{\mu\nu}q_L^i\sigma^{\mu\nu}q^j$.
Same as Table \ref{tb: scalar-yukawa-SppS/Tevatron}.
 }
\begin{tabular}[tb]{|c||c|c|c|c|c|}
\hline
\backslashbox{$i$}{$j$}  &  $\bar{b}$  &  $\bar{c}$  &  $\bar{s}$  & $\bar{u}$  &  $\bar{d}$ \\\hline\hline
$\bar{b}$& $2.8\times 10^{3}$ & $5.1$& $7.6$& \textcolor{green}{$\ast~4.3\times 10^{1}$} & $6.7$ \\\hline
$\bar{c}$& $5.1$& $5.6$& $4.2$& $8.4$& $4.1$\\\hline
$\bar{s}$& $7.6$& $4.2$& $8.8$& $1.8\times 10^{1}$ & $5.7$\\\hline
$\bar{u}$& \textcolor{green}{$\ast~4.3\times 10^{1}$} & $8.4$& $1.8\times 10^{1}$ & $2.6\times 10^{1}$ & $1.8\times 10^{1}$ \\\hline
$\bar{d}$& $6.7$& $4.1$& $5.7$& $1.8\times 10^{1}$ & $1.2\times 10^{1}$\\\hline\hline
$d$& $6.7$& $7.5$& $7.9$& $1.3\times 10^{1}$ & $1.2\times 10^{1}$ \\\hline
$u$&  \textcolor{green}{$\ast~4.3\times 10^{1}$} &  \textcolor{green}{$\ast~4.7\times 10^{1}$} &  \textcolor{green}{$\ast~5.0\times 10^{1}$} &  \textcolor{red}{$\bullet~6.8\times 10^{1}$} &  \textcolor{red}{$\bullet~6.7\times 10^{1}$} \\\hline
$s$& $7.6$& $8.7$& $8.8$& $1.2\times 10^{1}$ & $1.1\times 10^{1}$ \\\hline
$c$& $5.1$& $5.6$& $5.7$& $7.8$& $6.9$\\\hline
$b$& $2.8\times 10^{3}$ & $3.2\times 10^{3}$ & $3.2\times 10^{3}$ & $4.8\times 10^{3}$ & $4.1\times 10^{3}$ \\\hline
 \end{tabular}
\label{tb: vector-Pauli-SppS/Tevatron}
\end{center}
\end{table}
\begin{table}[h]
\begin{center}
\caption{($G_{\mu\nu}q^{\dag}\sigma^{\mu\nu}\psi$, \ref{sec: spinor-Pauli})
The required $N_\psi \Gamma(\psi\to qg)$ / 1 GeV for 4 pb at the Tevatron.
 }
\begin{tabular}[tb]{|c||c|c|c|c|c|}
\hline
  &  $\bar{b}$  &  $\bar{c}$  &  $\bar{s}$  & $\bar{u}$  &  $\bar{d}$ \\\hline\hline
$g$& $3.1\times 10^{1}$ & $1.8\times 10^{1}$ & $1.1\times 10^{1}$ & $6.7\times 10^{-1}$ & $1.6$\\\hline
 \end{tabular}
\label{tb: spinor-Pauli-decay width}
\end{center}

\begin{center}
\caption{($G_{\mu\nu}q^{\dag}\sigma^{\mu\nu}\psi$, \ref{sec:
 spinor-Pauli})
The required cut off $\Lambda$.
See the text for the definition of $\Lambda$.
}
\begin{tabular}[tb]{|c||c|c|c|c|c|}
\hline
  &  $\bar{b}$  &  $\bar{c}$  &  $\bar{s}$  & $\bar{u}$  &  $\bar{d}$ \\\hline\hline
$g$& $1.9\times 10^{2}$ & $2.4\times 10^{2}$ & $3.1\times 10^{2}$ & $1.3\times 10^{3}$ & $8.3\times 10^{2}$ \\\hline
 \end{tabular}
\label{tb: spinor-Pauli-cutoff}
\end{center}

\begin{center}
\caption{($G_{\mu\nu}q^{\dag}\sigma^{\mu\nu}\psi$, \ref{sec: spinor-Pauli})
The enhancement factor $R_{\rm LHC}$ for $G_{\mu\nu}q^{\dag}\sigma^{\mu\nu}\psi$.
}
\label{tb: spinor-Pauli-LHC7/Tevatron}
\begin{tabular}[tb]{|c||c|c|c|c|c|}
\hline
  &  $\bar{b}$  &  $\bar{c}$  &  $\bar{s}$  & $\bar{u}$  &  $\bar{d}$ \\\hline\hline
$g$& $9.5\times 10^{1}$ & $8.7\times 10^{1}$ & $7.9\times 10^{1}$ & $3.8\times 10^{1}$ & $4.8\times 10^{1}$ \\\hline
 \end{tabular}
\end{center}

\begin{center}
\caption{($G_{\mu\nu}q^{\dag}\sigma^{\mu\nu}\psi$, \ref{sec:
 spinor-Pauli})
The enhancement factor $R_{\rm SppS}$ for $G_{\mu\nu}q^{\dag}\sigma^{\mu\nu}\psi$.
}
\begin{tabular}[tb]{|c||c|c|c|c|c|}
\hline
  &  $\bar{b}$  &  $\bar{c}$  &  $\bar{s}$  & $\bar{u}$  &  $\bar{d}$ \\\hline\hline
$g$& \textcolor{green}{$\ast~2.1\times 10^{2}$} & \textcolor{red}{$\bullet~2.4\times 10^{2}$} & \textcolor{red}{$\bullet~2.4\times 10^{2}$} & \textcolor{red}{$\bullet~3.5\times 10^{2}$} & \textcolor{red}{$\bullet~3.2\times 10^{2}$} \\\hline
 \end{tabular}
\label{tb: spinor-Pauli-SppS/Tevatron}
\end{center}
\end{table}
\begin{table}[h]
\begin{center}
\caption{(Resonance, \ref{sec: resonance})
The required $N_Y\Gamma(Y\rightarrow ij)$ / 1 GeV for the Y cross
 section to be 4 pb at the Tevatron.
 }
\begin{tabular}[tb]{|c||c|c|c|c|c||c||}
\hline
\backslashbox{$i$}{$j$}  &  $\bar{b}$  &  $\bar{c}$  &  $\bar{s}$  & $\bar{u}$  &  $\bar{d}$ & $g$ \\\hline\hline
$\bar{b}$& $6.6\times 10^{1}$ & $3.8\times 10^{1}$ & $2.3\times 10^{1}$ & $1.5$& $3.5$& $4.2$\\\hline
$\bar{c}$& $3.8\times 10^{1}$ & $2.2\times 10^{1}$ & $1.3\times 10^{1}$ & $9.4\times 10^{-1}$ & $2.1$& $2.4$\\\hline
$\bar{s}$& $2.3\times 10^{1}$ & $1.3\times 10^{1}$ & $8$& $5.8\times 10^{-1}$ & $1.3$& $1.5$\\\hline
$\bar{u}$& $1.5$& $9.4\times 10^{-1}$ & $5.8\times 10^{-1}$ & $2.2\times 10^{-1}$ & $2.6\times 10^{-1}$ & $1.1\times 10^{-1}$ \\\hline
$\bar{d}$& $3.5$& $2.1$& $1.3$& $2.6\times 10^{-1}$ & $4.1\times 10^{-1}$ & $2.3\times 10^{-1}$ \\\hline\hline
$g$& $4.2$& $2.4$& $1.5$& $1.1\times 10^{-1}$ & $2.3\times 10^{-1}$ & $2.7\times 10^{-1}$ \\\hline\hline
$d$& $3.5$& $2.1$& $1.3$& $9.3\times 10^{-2}$ & $1.8\times 10^{-1}$ & $2.3\times 10^{-1}$ \\\hline
$u$& $1.5$& $9.4\times 10^{-1}$ & $5.8\times 10^{-1}$ & $4.9\times 10^{-2}$ & $9.3\times 10^{-2}$ & $1.1\times 10^{-1}$ \\\hline
$s$& $2.3\times 10^{1}$ & $1.3\times 10^{1}$ & $8$& $5.8\times 10^{-1}$ & $1.3$& $1.5$\\\hline
$c$& $3.8\times 10^{1}$ & $2.2\times 10^{1}$ & $1.3\times 10^{1}$ & $9.4\times 10^{-1}$ & $2.1$& $2.4$\\\hline
$b$& $6.6\times 10^{1}$ & $3.8\times 10^{1}$ & $2.3\times 10^{1}$ & $1.5$& $3.5$& $4.2$\\\hline
 \end{tabular}
\label{tb: resonance-decaywidth}
\vspace{1cm}
\caption{(Resonance, \ref{sec: resonance})
The enhancement factor $R_{\rm LHC}$. The elements marked $\circ$ and written in blue letters are NOT excluded by
 the ATLAS result.}
\begin{tabular}[tb]{|c||c|c|c|c|c||c||}
\hline
  &  $\bar{b}$  &  $\bar{c}$  &  $\bar{s}$  & $\bar{u}$  &  $\bar{d}$ & $g$ \\\hline\hline
$\bar{b}$& $7.4\times 10^{1}$ & $6.7\times 10^{1}$ & $5.8\times 10^{1}$ & $2.2\times 10^{1}$ & $2.9\times 10^{1}$ & $6.1\times 10^{1}$ \\\hline
$\bar{c}$& $6.7\times 10^{1}$ & $6.1\times 10^{1}$ & $5.2\times 10^{1}$ & $1.9\times 10^{1}$ & $2.6\times 10^{1}$ & $5.5\times 10^{1}$ \\\hline
$\bar{s}$& $5.8\times 10^{1}$ & $5.2\times 10^{1}$ & $4.4\times 10^{1}$ & $1.5\times 10^{1}$ & $2.1\times 10^{1}$ & $4.7\times 10^{1}$ \\\hline
$\bar{u}$& $2.2\times 10^{1}$ & $1.9\times 10^{1}$ & $1.5\times 10^{1}$ & $1.7\times 10^{1}$ & $1.5\times 10^{1}$ & $1.8\times 10^{1}$ \\\hline
$\bar{d}$& $2.9\times 10^{1}$ & $2.6\times 10^{1}$ & $2.1\times 10^{1}$ & $1.5\times 10^{1}$ & $1.6\times 10^{1}$ & $2.4\times 10^{1}$ \\\hline\hline
$g$& $6.1\times 10^{1}$ & $5.5\times 10^{1}$ & $4.7\times 10^{1}$ & $1.8\times 10^{1}$ & $2.4\times 10^{1}$ & $5.0\times 10^{1}$ \\\hline\hline
$d$& $2.9\times 10^{1}$ & $2.6\times 10^{1}$ & $2.1\times 10^{1}$ &\textcolor{blue}{$\circ~4.1$}& \textcolor{blue}{$\circ~6.2$}&$2.4\times 10^{1}$ \\\hline
$u$& $2.2\times 10^{1}$ & $1.9\times 10^{1}$ & $1.5\times 10^{1}$ & \textcolor{blue}{$\circ~2.6$}& \textcolor{blue}{$\circ~4.1$}& $1.8\times 10^{1}$ \\\hline
$s$& $5.8\times 10^{1}$ & $5.2\times 10^{1}$ & $4.4\times 10^{1}$ & $1.5\times 10^{1}$ & $2.1\times 10^{1}$ & $4.7\times 10^{1}$\\\hline
$c$& $6.7\times 10^{1}$ & $6.1\times 10^{1}$ & $5.2\times 10^{1}$ & $1.9\times 10^{1}$ & $2.6\times 10^{1}$ & $5.5\times 10^{1}$ \\\hline
$b$& $7.4\times 10^{1}$ & $6.7\times 10^{1}$ & $5.8\times 10^{1}$ & $2.2\times 10^{1}$ & $2.9\times 10^{1}$ & $6.1\times 10^{1}$ \\\hline
 \end{tabular}
\label{tb: resonance-LHC7/Tevatron}
\end{center}
\end{table}

\clearpage
\section{Possible improvement of the search at LHC}
\label{sec:LHCimprovement}
In the previous sections, we find that some operators are insensitive to the current search
at the LHC 7 TeV run and the ${\rm Sp\bar{p}S}$.
We list such ceses here;
\begin{eqnarray}
 \phi d_R^{\dag}d_L,~\phi d_R^{\dag}u_L,~V^{\mu}
  u_L^{\dag}\bar{\sigma}_{\mu}d_L,~V^{\mu}
  u_L^{\dag}\bar{\sigma}_{\mu}d_L,~V^{\mu\nu}
  u_R^{\dag}\bar{\sigma}_{\mu\nu}d_L,~V^{\mu\nu}
  d_R^{\dag}\bar{\sigma}_{\mu\nu}d_L \nonumber \\
 {\rm Resonant~production,~the~initial~partons~are}~d\bar{u},~u\bar{u},~d\bar{d}
\label{eq:allowed}
\end{eqnarray}
In this section, we study a possible improvement of the
search at the LHC 7 TeV run. We apply the following selection cuts in addition to the ones
used in Ref.~\cite{ATLAS_wjj}.
\begin{enumerate}
 \item The leading two jets have $p_T$ larger than 100 and 50 GeV,
 \item There is only one lepton which is require to have $p_T$ larger
       than 90 GeV,
 \item $\slashed{E}_T$ is larger than 90 GeV,
 \item 140 GeV $<~M_{jj}~<$ 160 GeV.
\end{enumerate}
Here $p_T$, $\slashed{E}_T$ and $M_{jj}$ are the transverse momentum,
the transverse missing energy and the invariant mass of the leading two
jets, respectively. The number of the remaining background events after applying
 all the cuts are summarized in Table.~\ref{table:LHCbg}.
We also summarize the acceptance of the signal for some representative
cases in Table.~\ref{table:LHCsig}.
Unfortunately, the acceptances of the signals strongly depend on the operators
 which produce the $Wjj$ signal with the severe cut we
 apply. Especially, in the cases of the resonant production, the
 acceptances are too small that the signals are not observed in the
 LHC. Such low acceptances is expected because the energies of $W$ and $X$ are
 restricted by the mass of the parent particle $Y$. We must
 use more sophisticated techniques to search for the resonant
 production, which is model-dependent and beyond the scope of this
 article.
In the case of the operator $V_{\mu\nu}u_R^{\dag} \sigma^{\mu\nu}d_{L}$,
the acceptance is rather high. This is because the operator is the
higher dimensional one and the cross section do not drop down at high
energy.
To estimate the expected constrains on the operators, it is conservative
to use the acceptance for the operator $\phi u_L^{\dag}d_R$, which is renormalizable.
For the non-resonant production, with the use of the conservative acceptance,
the expected 95\% C.L. constrains on $R_{{\rm LHC}}$ is $R_{{\rm LHC}}<
1.4$ with 5 fb$^{-1}$ data. Here we assign the same degree of systematic error
as the one in Ref.~\cite{ATLAS_wjj}. This expected constraint covers all the
operators we list in Sec.~\ref{sec:nonresonant}. Since we use the fast
simulation, this expected constraint is just a rough estimate. Especially, the
systematic error should be evaluated based on the actual experiments. 

\begin{table}
\begin{center}
\caption{
The expected number of the events after applying all the selection cuts listed in
 Sec.~\ref{sec:LHCimprovement} in addition to the ones applied in Ref.~\cite{ATLAS_wjj} at the LHV 7 TeV run.
}
\begin{tabular}{|c||c|}
\hline
& Number of the events $/$fb$^{-1}$\\
\hline
$W+$ jets & 50.3 \\
$t\bar{t}+$ jets & 52.5 \\
\hline
\end{tabular}
\label{table:LHCbg}
\end{center}
\end{table}

\begin{table}
\begin{center}
\caption{
The acceptance of the signal after applying all the selection cuts listed in
 Sec.~\ref{sec:LHCimprovement} in addition to the ones applied in Ref.~\cite{ATLAS_wjj} at the LHV 7 TeV run.
}
\begin{tabular}{|c||c|}
\hline
& Acceptances\\
\hline
$\phi u_L^{\dag}d_R$ & 0.013  \\
$V_{\mu\nu}u_R^{\dag} \sigma^{\mu\nu}d_{L}$ & 0.035\\
Resonant production, initial parton are $\bar{d}u$ & 0 \\
\hline
\end{tabular}
\label{table:LHCsig}
\end{center}
\end{table}
 
\clearpage

\section{Conclusion and Discussion}
In this paper, we discuss testability of the CDF $Wjj$ anomaly at the
LHC.
We comprehensively study models which can realize $Wjj$ signal observed
by the CDF collaboration.
We have found that the cross section at the LHC mainly depends on what
partons in $p$ and $\bar{p}$
produce $W$and $X$ and that
almost all the models are inconsistent with the result of the LHC,
unless only valence quarks contribute the new process. We list the
allowed cases in Eq.~(\ref{eq:allowed}).
We also study the possible improvement of the search at LHC. We have
found that by using the more severe cuts than the one used in
\cite{ATLAS_wjj}, it might be possible to discover/exclude the $Wjj$
signal for any models except for the case of resonant production.
As mentioned in footnote 1, some models are not within our
study.
However, even in such a case, the cross section at the LHC is expected
to be ${\cal O}(1) \times \sigma_{\rm Tevatron}$
in the case of production by valence quarks and ${\cal O}(10-100) \times
\sigma_{\rm Tevatron}$ for sea quarks or gluons case.
In the latter case, the ATLAS result gives strong constraint.

\section*{Acknowledgement}
This work was supported by the World Premier International Research Center Initiative (WPI Initiative), MEXT, Japan. The work of RS and SS is supported in part by JSPS Research Fellowships for Young Scientists.


\begin{thebibliography}{99}


\bibitem{Aaltonen:2011mk}
  T.~Aaltonen {\it et al.}  [CDF Collaboration],
  Phys.\ Rev.\ Lett.\  {\bf 106} (2011) 171801
  [arXiv:1104.0699 [hep-ex]].





\bibitem{Buckley:2011vc}
  M.~R.~Buckley, D.~Hooper, J.~Kopp, E.~Neil,
  [arXiv:1103.6035 [hep-ph]];
  F.~Yu,
  Phys.\ Rev.\  {\bf D83}, 094028 (2011).
  [arXiv:1104.0243 [hep-ph]];
  X.~-P.~Wang, Y.~-K.~Wang, B.~Xiao, J.~Xu, S.~-h.~Zhu,
  [arXiv:1104.1161 [hep-ph]];
  K.~Cheung, J.~Song,
  [arXiv:1104.1375 [hep-ph]];
  X.~-P.~Wang, Y.~-K.~Wang, B.~Xiao, J.~Xu, S.~-h.~Zhu,
  [arXiv:1104.1917 [hep-ph]];
  L.~A.~Anchordoqui, H.~Goldberg, X.~Huang, D.~Lust, T.~R.~Taylor,
  [arXiv:1104.2302 [hep-ph]];
  S.~Jung, A.~Pierce, J.~D.~Wells,
  [arXiv:1104.3139 [hep-ph]];
  P.~Ko, Y.~Omura, C.~Yu,
  [arXiv:1104.4066 [hep-ph]];
  P.~J.~Fox, J.~Liu, D.~Tucker-Smith, N.~Weiner,
  [arXiv:1104.4127 [hep-ph]];
  D.~-W.~Jung, P.~Ko, J.~S.~Lee,
  [arXiv:1104.4443 [hep-ph]];
  S.~Chang, K.~Y.~Lee, J.~Song,
  [arXiv:1104.4560 [hep-ph]];
  X.~Huang,
  [arXiv:1104.5389 [hep-ph]];
  Z.~Liu, P.~Nath, G.~Peim,
  [arXiv:1105.4371 [hep-ph]];
  J.~L.~Hewett, T.~G.~Rizzo,
  [arXiv:1106.0294 [hep-ph]];
  A.~E.~Faraggi, V.~M.~Mehta,
  [arXiv:1106.5422 [hep-ph]];
  L.~Vecchi,
  [arXiv:1107.2933 [hep-ph]].


\bibitem{Kilic:2011sr}
  C.~Kilic, S.~Thomas,
  [arXiv:1104.1002 [hep-ph]];
  A.~E.~Nelson, T.~Okui, T.~S.~Roy,
  [arXiv:1104.2030 [hep-ph]];
  B.~A.~Dobrescu, G.~Z.~Krnjaic,
  [arXiv:1104.2893 [hep-ph]];
  Q.~-H.~Cao, M.~Carena, S.~Gori, A.~Menon, P.~Schwaller, C.~E.~M.~Wagner, L.~-T.~Wang,
  [arXiv:1104.4776 [hep-ph]];
  B.~Dutta, S.~Khalil, Y.~Mimura, Q.~Shafi,
  [arXiv:1104.5209 [hep-ph]];
  L.~M.~Carpenter, S.~Mantry,
  [arXiv:1104.5528 [hep-ph]];
  G.~Segre, B.~Kayser,
  [arXiv:1105.1808 [hep-ph]];
  T.~Enkhbat, X.~-G.~He, Y.~Mimura, H.~Yokoya,
  [arXiv:1105.2699 [hep-ph]];
  C.~-H.~Chen, C.~-W.~Chiang, T.~Nomura, Y.~Fusheng,
  [arXiv:1105.2870 [hep-ph]];
  A.~Alves, E.~R.~Barreto, A.~G.~Dias,
  [arXiv:1105.4849 [hep-ph]];
  J.~Fan, D.~Krohn, P.~Langacker, I.~Yavin,
  [arXiv:1106.1682 [hep-ph]];
  J.~Evans, B.~Feldstein, W.~Klemm, H.~Murayama, T.~T.~Yanagida,
  [arXiv:1106.1734 [hep-ph]];
  J.~F.~Gunion,
  [arXiv:1106.3308 [hep-ph]];
  D.~K.~Ghosh, M.~Maity, S.~Roy,
  [arXiv:1107.0649 [hep-ph]].
\bibitem{Isidori:2011dp}
  G.~Isidori, J.~F.~Kamenik,
  Phys.\ Lett.\  {\bf B700}, 145-149 (2011).
  [arXiv:1103.0016 [hep-ph]];
  R.~Sato, S.~Shirai, K.~Yonekura,
  Phys.\ Lett.\  {\bf B700}, 122-125 (2011).
  [arXiv:1104.2014 [hep-ph]].

\bibitem{Anchordoqui:2010zs}
  L.~A.~Anchordoqui, W.~-Z.~Feng, H.~Goldberg, X.~Huang, T.~R.~Taylor,
  Phys.\ Rev.\  {\bf D83}, 106006 (2011).
  [arXiv:1012.3466 [hep-ph]];
  E.~J.~Eichten, K.~Lane, A.~Martin,
  [arXiv:1104.0976 [hep-ph]];
  H.~B.~Nielsen,
  [arXiv:1104.4642 [hep-ph]];
  K.~S.~Babu, M.~Frank, S.~K.~Rai,
  [arXiv:1104.4782 [hep-ph]];
  R.~Harnik, G.~D.~Kribs, A.~Martin,
  [arXiv:1106.2569 [hep-ph]];
  Y.~Cui, Z.~Han, M.~D.~Schwartz,
  [arXiv:1106.3086 [hep-ph]].
  R.~Fok, G.~D.~Kribs,
  [arXiv:1106.3101 [hep-ph]].

\bibitem{He:2011ss} 
  X.~-G.~He and B.~-Q.~Ma,
  Eur.\ Phys.\ J.\ A {\bf 47}, 152 (2011)
  [arXiv:1104.1894 [hep-ph]];
  Z.~Sullivan and A.~Menon,
  Phys.\ Rev.\ D {\bf 83}, 091504 (2011)
  [arXiv:1104.3790 [hep-ph]];
  T.~Plehn and M.~Takeuchi,
  J.\ Phys.\ G G {\bf 38}, 095006 (2011)
  [arXiv:1104.4087 [hep-ph]].


\bibitem{Abazov:2011af}
  V.~M.~Abazov {\it et al.}  [D0 Collaboration],
  arXiv:1106.1921 [hep-ex].

\bibitem{Eichten:2011xd}
  E.~Eichten, K.~Lane, A.~Martin,
  [arXiv:1107.4075 [hep-ph]].

\bibitem{CDF_lvjj}
A.~Annovi, P.~Catastini, V.~Cavaliere, L.~Ristori,
``Kinematic Distribution of events in the $115 < M_{JJ} < 175$ GeV region."\\
{\tt http://www-cdf.fnal.gov/physics/ewk/2011/wjj/kinematics.html}

\bibitem{ATLAS_wjj}
The ATLAS Collaboration,
``Invariant mass distribution of jet pairs produced in association with a
leptonically decaying $W$ boson using 1.02 fb$^{-1}$ of ATLAS data", ATLAS-CONF-2011-097,\\
{\tt http://cdsweb.cern.ch/record/1369206/files/ATLAS-CONF-2011-097.pdf}

\bibitem{Alwall:2007st}
  J.~Alwall, P.~Demin, S.~de Visscher, R.~Frederix, M.~Herquet, F.~Maltoni, T.~Plehn, D.~L.~Rainwater {\it et al.},
  JHEP {\bf 0709}, 028 (2007).
  [arXiv:0706.2334 [hep-ph]].

\bibitem{Moneta:2010pm}
  L.~Moneta, K.~Belasco, K.~Cranmer, A.~Lazzaro, D.~Piparo, G.~Schott, W.~Verkerke, M.~Wolf {\it et al.},
  PoS {\bf ACAT2010}, 057 (2010).
  [arXiv:1009.1003 [physics.data-an]].



\bibitem{Alitti:1993pn}
  J.~Alitti {\it et al.} [ UA2 Collaboration ],
  Nucl.\ Phys.\  {\bf B400}, 3-24 (1993).
  
\bibitem{Stump:2003yu}
  D.~Stump, J.~Huston, J.~Pumplin, W.~-K.~Tung, H.~L.~Lai,
S.~Kuhlmann, J.~F.~Owens,
  JHEP {\bf 0310}, 046 (2003).
  [hep-ph/0303013].


\end{thebibliography}
\end{document}